\documentclass[a4paper]{article}

\usepackage{INTERSPEECH2021}
\usepackage{multirow}
\usepackage{makecell}
\usepackage{graphicx,color}
\usepackage{comment}
\usepackage{amsmath,amssymb} % define this before the line numbering.
\usepackage{bbm}
\usepackage{color}
\usepackage{xcolor}
\usepackage{adjustbox}
\usepackage{enumitem}
\usepackage{subfigure}
\usepackage[linesnumbered, ruled, vlined, algo2e]{algorithm2e}
\usepackage{array}
\usepackage{caption}
\usepackage{soul}
\usepackage{makecell}
\usepackage{soul}

\title{Exploring Targeted Universal Adversarial Perturbations \\
to End-to-end ASR Models} 

\name{Zhiyun Lu, Wei Han, Yu Zhang, Liangliang Cao}

\address{Google Inc., USA}
\email{\{zhiyunlu,weihan,ngyuzh,llcao\}@google.com}

\usepackage{amssymb}
\usepackage{amsmath,amsfonts}
\usepackage{amsopn,bm,mathtools}

\usepackage{nicefrac}       % compact symbols for 1/2, etc.
\newcommand{\vct}[1]{\boldsymbol{#1}} % vector
\newcommand{\ProbOpr}[1]{\mathbb{#1}}
\newcommand{\expect}[2]{%
\ifthenelse{\equal{#2}{}}{\ProbOpr{E}_{#1}}
{\ifthenelse{\equal{#1}{}}{\ProbOpr{E}\left[#2\right]}{\ProbOpr{E}_{#1}\left[#2\right]}}} % Expectation: syntax: E{1}{2} = E_1[2], E{}{2}=E[2], E{1}{} = E_1
\newcommand{\var}[2]{%
\ifthenelse{\equal{#2}{}}{\ProbOpr{VAR}_{#1}}
{\ifthenelse{\equal{#1}{}}{\ProbOpr{VAR}\left[#2\right]}{\ProbOpr{VAR}_{#1}\left[#2\right]}}} % Expectation: syntax: V{1}{2} = V_1[2], V{}{2}=V[2], V{1}{} = V_1
\newcommand{\vx}{{\vct{x}}}

\newcommand{\sT}{\mathcal{T}}

\newcommand{\vdelta}{\vct{\delta}}

\usepackage{xspace}
\makeatletter
\DeclareRobustCommand\onedot{\futurelet\@let@token\@onedot}
\def\@onedot{\ifx\@let@token.\else.\null\fi\xspace}
\def\eg{\emph{e.g}\onedot} 
\def\ie{\emph{i.e}\onedot}

\makeatother

\usepackage{mathtools}

\usepackage[ruled,vlined]{algorithm2e}

\usepackage{listings}

\newcommand{\eat}[1]{}

\begin{document}

\maketitle
\begin{abstract}
Although end-to-end automatic speech recognition (e2e ASR) models are widely deployed in many applications, there have been very few studies to understand models' robustness against adversarial perturbations. In this paper, we explore whether a targeted universal perturbation vector exists for e2e ASR models. Our goal is to find perturbations that can mislead the models to predict the given targeted transcript such as ``thank you" or empty string on any input utterance. We study two different attacks, namely additive and prepending perturbations, and their performances on the state-of-the-art LAS, CTC and RNN-T models. We find that LAS is the most vulnerable to perturbations among the three models. RNN-T is more robust against additive perturbations, especially on long utterances. And CTC is robust against both additive and prepending perturbations. 
To attack RNN-T, we find prepending perturbation is more effective than the additive perturbation, and can mislead the models to predict the same short target on utterances of arbitrary length. 
\end{abstract}
% Adversarial perturbation is extensively studied in computer vision and deep learning communities, but not studied thoroughly in automatic speech recognition. 
% (i.e., audio-agnostic)
     %intelligent personal assistants, 
          %it becomes increasingly important 
%\llcao{add ``many"? coz we cannot attack CTC }
% LAS is the most vulnerable architecture among the three under audio-agnostic attack because of the attention mechanism, which allows the model to skip speech frames. 
% The culprit is the attention mechanism which can be easily fooled to attend to the noisy speech and skip the rest of the utterance. 
% \llcao{note for zhiyunlu: previously you first say the universal attacks exist for all e2e models. and then explain different conclusions for ctc/rnnt/las. now I changed to ``explore", so the new sentence will not conflict or duplicate with the following sentences.  }
%We study both additive and prepending perturbations. We benchmark the performance of the attack on the state-of-the-art LAS, RNN-T and CTC models. 
% \llcao{not precise. any idea to say this better? otherwise we can use the current one}
\section{Introduction} \label{sec:intro}

Adversarial example~\cite{szegedy2013intriguing,goodfellow2014explaining,carlini2017towards} is a carefully crafted input that fools the network into wrong predictions by applying a worst-case perturbation to a test sample. 
Adversarial perturbations are extensively studied in vision~\cite{szegedy2013intriguing,moosavi2017universal} and language~\cite{alzantot2018generating,jia2017adversarial}. In audio domain, ~\cite{abdoli2019universal, vadillo2019universal, xie2020real} studies adversarial examples for sound/speech classification and speaker identification problems, where the output is a single class instead of a label sequence. 

Adversarial example for automatic speech recognition (ASR) models~\cite{carlini2016hidden,abdullah2019practical,alzantot2018did,carlini2018audio,qin2019imperceptible, abdullah2020sok} is substantially different from those in the image domain~\cite{abdullah2020sok}, due to the sequential predictions. For example, an \emph{untargeted} adversarial attacks can easily yield high word error rate (WER) to ASR by introducing spelling errors~\cite{qin2019imperceptible}, or by inserting random words to the transcript, while preserving much of the ground truth's text or semantics. This is very different from the attack to image classification models, where a high error rate guarantees that users can not relate the mis-classification with the ground truth. To this end, we study \emph{targeted} attack in this work. In addition to high WER, the ASR model should predict a specific mis-transcription target. Targeted attack is much more challenging~\cite{carlini2018audio,cisse2017houdini} than untargeted attack.

\begin {figure}[tbp]
% \captionsetup{farskip=0pt}% <--- no gap at the top
\hspace{-0.1 in}
% \vspace{-0.1 in}
\centering
    \includegraphics[width= 0.48\textwidth]{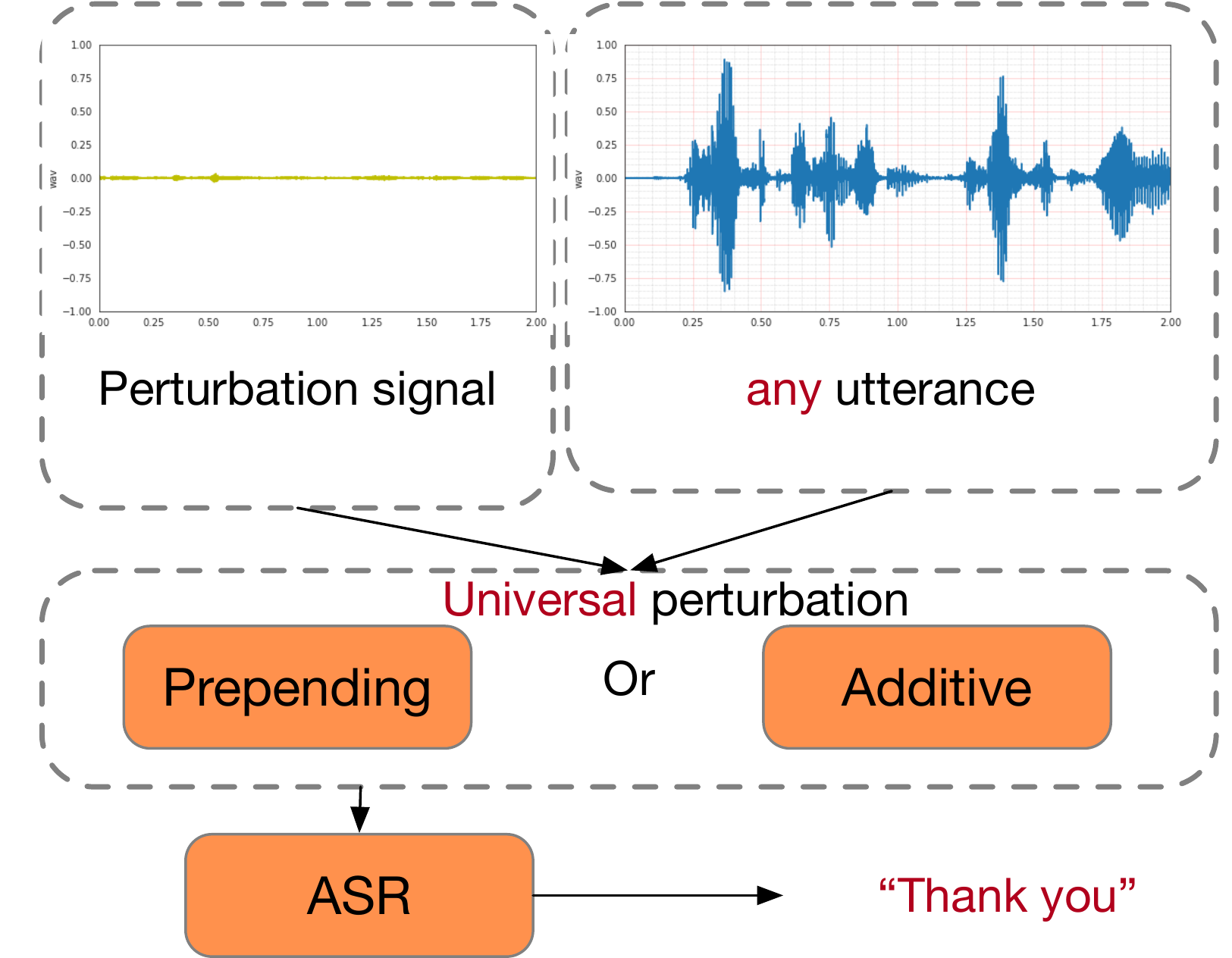}
% \vspace{-0.2 in}
\caption{System overview: We apply a universal perturbation to the utterance,which causes the ASR model to output the same transcript, \eg ``thank you'', for \textbf{any} utterance. The input utterance can be of different lengths and arbitrary content. 
We study both prepending and additive perturbations.}\label{fig:diagram}
\vspace{-0.15 in}
\end{figure}

A large portion of previous works focus on input-dependent perturbations~\cite{carlini2018audio, qin2019imperceptible}, where for each utterance we solve a separate optimization to compute the perturbation. On the other hand, a \emph{universal} (audio-agnostic) perturbation is one perturbation vector, learnt from training set, which can generalize and cause \emph{any} audio to be mis-transcribed with high probability. Universal perturbation can attack unseen utterances in real-time without any new optimization. As it is more efficient and practical, we focus on universal perturbations. Figure~\ref{fig:diagram} illustrates targeted universal perturbation: for any utterance, a universal perturbation is applied to the utterance. The ASR model is fooled to predict a specific transcript, e.g. ``thank you'', regardless of the input. Both prepending and additive perturbations are studied in the experiments.

One challenge of adversarial attack to ASR models lies in the fast evolution of models. In the past five years, various e2e ASR models have been deployed on both servers and mobile devices, including CTC~\cite{graves2006connectionist}, LAS~\cite{chan2015listen} and RNN-T~\cite{graves2012sequence}, and RNN-T models have pushed the-state-of-the-art~\cite{gulati2020conformer} on the Librispeech dataset~\cite{panayotov2015librispeech}. In contrast, most of the studies on ASR adversarial attacks are limited to a single type of model. For example,~\cite{qin2019imperceptible} focuses on LAS model, and~\cite{carlini2018audio,neekhara2019universal} on CTC. 

This paper aims to conduct a comprehensive study on the targeted universal adversarial perturbations against different ASR models.
There are two main discoveries from our research. Firstly, we show that universal perturbation exists for e2e ASR models on some given targets. The learnt universal perturbation can fool the state-of-the-art ASR models to generate the same mis-transcription target on Librispeech~\cite{panayotov2015librispeech} test sets. On short sentences composed of frequent words, the success rate of the perturbation can be over 99\%. For long targets, the success rate will be much lower.
Secondly, we find different e2e models behave differently against the perturbations. We compared two perturbations, additive perturbation, studied in~\cite{qin2019imperceptible,neekhara2019universal}, and prepending perturbation, a new type of perturbation for speech data. 
We find LAS model is vulnerable to both additive and prepending perturbations, while RNN-T is more robust against additive perturbation but becomes vulnerable to prepending perturbations. For a state-of-the-art RNN-T model trained from Librispeech, we can append a 4 seconds long audio clip to utterances of arbitrary lengths, and the ASR model would always generate the same wrong predictions. Lastly, CTC is robust against both perturbations. Note that our work is not concerned about the imperceptibility of the perturbation, as motivated in~\cite{brown2017adversarial}.

This paper is organized as follows: \S~\ref{sec:method} introduces the learning framework of the target universal perturbation. In \S~\ref{sec:setup} we describe the data as well as the evaluation metrics of the experimental study. Extensive experiments on different e2e models are discussed in \S~\ref{sec:exp}, and \S~\ref{sec:conclude} concludes the paper. 
%!TEX root = main.tex
\section{Method} \label{sec:method}
We formally define the targeted universal perturbation problem in \S~\ref{sec:problem}, and discuss the perturbation function in \S~\ref{sec:perturb}. Lastly we provide the architecture details of the e2e models used in the empirical study in \S~\ref{sec:model}. 

\begingroup

\subsection{Targeted universal adversarial perturbation}\label{sec:problem}
ASR is a model the predicts a text sequence given an input utterance. Define $\vx = [x_1, \ldots, x_T]$ as the input sequence, and $\vdelta = [\delta_1, \ldots, \delta_S]$ as the perturbation sequence, where $S$ and $T$ are the sequence lengths. We denote the perturbed input as $\sT(\vdelta, \vx)$, which applies perturbation $\vdelta$ to $\vx$ through operation $\sT$. We will discuss the form of $\sT ({\vdelta}, \vx)$ in next section. 
The text sequence is denoted as $y$. For an input-output pair $(\vx, y)$, we denote the loss of the ASR model as $\ell(\vx, y)$. Here we ignore the dependency on model parameters for brevity since the parameters are fixed in the attack problem. Note $\ell$ is different for different models: cross-entropy for attention model, RNN-T loss for RNN-T model, and CTC loss for CTC model.

For targeted perturbation, we specify a mis-transcription $y'$. Our goal is to find a universal perturbation $\vdelta$ that can attack the ASR system to output $y'$ for any utterances $\vx$, drawn from some distribution or dataset $\mathcal{D}$. Mathematically a sufficient condition is that $\ell(\sT({\vdelta},\vx), y')$ is small across all samples $\forall \vx \in \mathcal{D}$. Therefore the targeted audio-agnostic adversarial attack can be formulated as the following optimization problem
\begin{align}
\min_{\vdelta} \sum_{x\in \mathcal{D}} \ell(\sT(\vdelta, \vx), y'), \quad \text{s.t. } \|\vdelta\|_{\infty} < \epsilon, \label{eq:targetloss}
\end{align}
where $\epsilon$ is the maximum allowed $L_\infty$ norm of the perturbation, following~\cite{qin2019imperceptible,carlini2018audio,neekhara2019universal}. Since we are not concerned about imperceptibility in this work, we choose $\epsilon$ to be $2^{15}$, the signal range for a 16-bit PCM format audio, for most of the experiments. 
While more sophisticated loss functions specific to architectures are possible, like the one discussed in~\cite{carlini2018audio}, we try to keep a simple and unified framework that works for all models in this study. We solve the minimization problem by stochastic gradient descent.
% As argued in~\cite{carlini2018audio}, FGSM might not be appropriate for audio data due to the inherent non-linearity introduced in spectrogram computations.

\subsection{Perturbation $\sT(\vdelta, \vx)$}\label{sec:perturb}
While perturbation on the frequency domain is possible, we focus on generating adversarial perturbations in the audio-domain.

\noindent \textbf{Additive perturbation.} This is the perturbation studied in previous works~\cite{abdoli2019universal,carlini2018audio,qin2019imperceptible}. It is a direct adaptation of image perturbation~\cite{moosavi2017universal} to audio domain. $\vdelta$ is truncated or padded with $0$ to the same length of $\vx$ and then added to $\vx$. Formally,

\begin{equation}
\sT({\vdelta}, \vx) = [x_1 + \bar{\delta}_1, x_2 + \bar{\delta}_2, \ldots, x_T + \bar{\delta}_T],
\end{equation}
where $\bar{\delta}_t = \delta_t$ when $t\leq S$, and $0$ otherwise.

\noindent \textbf{Prepending perturbation.} %\llcao{Prepending or prepended?}
We propose a new adversarial perturbation, observing that speech is of sequential nature and that the input is of variable length. Prepending perturbation is defined as
\begin{equation}
\sT(\vdelta, \vx) = [\delta_1, \ldots, \delta_S, x_1, \ldots, x_T].
\end{equation}
where $\vdelta$ is concatenated before $\vx$.
Note that prepending perturbation does not contaminate the utterance at all except that the model runs for a few steps with $\vdelta$ as input before seeing $\vx$. This perturbation attacks e2e models' causal stability, while additive perturbation measures the networks' additive stability~\cite{szegedy2013intriguing}.

\subsection{Models}\label{sec:model}
We focus on the e2e ASR models, and study three most popular architectures: RNN-T~\cite{graves2012sequence}, attention model~\cite{chan2015listen}, and CTC~\cite{graves2006connectionist}. For RNN-T model, we use the state-of-the-art Conformer (small) model introduced in~\cite{gulati2020conformer}. It has an encoder of 16 layers conformer blocks of 144 dimensions, and a single LSTM layer decoder with 1024 hidden units and a 640 units projection layer. The joint network has 640 units. For attention model, we use the same encoder and decoder as in the RNN-T model, and with a multi-head attention of 4 heads and hidden size 288. For CTC model, the encoder is the same as RNN-T, but the decoder is a simple projection layer of dimension 1024. 
For all three models, we extracted 80-channel filterbanks features computed from a 25ms window with a stride of 10ms. The tokenizer is a 1k word-piece model built from LibriSpeech960h.
We use SpecAugment~\cite{park2019specaugment} during training. Please refer to~\cite{gulati2020conformer} for more details about the training hyper-parameters. 
All models and the attack are implemented and trained with Lingvo toolkit~\cite{shen2019lingvo}. The number of parameters in the model and the WERs on Librispeech test-clean and test-other sets are given in Table~\ref{tab:wer}. To provide a baseline for the prepending perturbation, we report the WER of the test sets with 4 seconds of silence (numerical 0) prepended to each utterance in the last column. 

\begin{table}[t]
    \centering
    \caption{Number of parameters and WERs on Librispeech test sets of e2e ASR models. The last column shows the WER of the test sets with 4 seconds of silence prepended to each utterance. It serves as an baseline for the prepending perturbation.}    \label{tab:wer}
    \begin{tabular}{|c|c|c|c|}
    \hline
    model & \# params (M) & \makecell[c]{WER \\ clean / other} & 
    \makecell[c]{WER pp-4s \\ clean / other} \\ \hline
    LAS & 10.3 & 4.9 / 7.9  &  11.9 / 13.8 \\
    RNN-T & 10.4 & 2.6 / 6.4 &   4.5 / 9.3 \\ 
     CTC & 8.9 & 3.6 / 8.7  & 3.6 / 8.8 \\ \hline
    \end{tabular}
    \vspace{-0.2 in}
\end{table}
\endgroup 
%!TEX root = main.tex
\section{Experimental Setup}\label{sec:setup}
\subsection{Dataset} 
LibriSpeech960h dataset~\cite{panayotov2015librispeech} is used to train the perturbations as well as the e2e ASR models (see \S~\ref{sec:model}) in our experiments. It is a corpus of 16KHz English speech from audiobooks. The training set has 281,241 utterances. We report the attack results on Librispeech test-clean and test-other sets, which contain 2620 and 2939 utterances respectively. 

Throughout our experiments, the perturbation is of 4 seconds long, for both additive and prepending types. For the attack optimization\footnote{We did not encounter small gradient issue after properly tuning the batch size, thus the vanilla SGD is preferred over the Fast Gradient Sign Method~\cite{goodfellow2014explaining} variant for simplicity.}, we use Adam~\cite{kingma2014adam} optimizer. The learning rate is 1.0 for the first 50k steps, and exponentially decays after 50k steps. The batch size is 1024. We set the the perturbation max norm constraint $\epsilon$ in Eq.~\eqref{eq:targetloss} to be $2^{15}$, which is the signal range for a 16-bit PCM format audio, if not otherwise specified. Our main focus is to study the feasibility of universal perturbation, similar as in~\cite{brown2017adversarial}, and imperceptibility is not our concern.

\subsection{Evaluation metrics}
To evaluate the performance of the targeted attack, following existing works~\cite{carlini2018audio, qin2019imperceptible} we report, 
(1) Success rate (sentence-level accuracy):  ${N_s}/{N}$ where $N_s$ is the number of audios that are transcribed as $y'$, and $N$ is the total number of audios in the test set;
(2) Distortion: We quantify the relative
loudness of the perturbation $\vdelta$ with respect to the original
audio $\vx$ in decibels (dB): $D(\vdelta, \vx) = \text{dB}(\vdelta) - \text{dB}(\vx)$ where $
\text{dB}(\vx) = 20\log_{10}(\max_t(x_t))$. The success rate is the higher the better, while the distortion is the lower the better.
%!TEX root = main.tex
\section{Experimental Results} \label{sec:exp}
There are three variables that determine a targeted attack: the perturbation (see \S~\ref{sec:perturb}), the model architecture (see \S~\ref{sec:model}), and the mis-transcription target $y'$.
In the experiment study, we try to answer the following questions: (a) which ASR model is vulnerable to targeted attacks; (b) what perturbation $\sT$ is effective to attack ASR models; (c) what mis-transcription $y'$ is more likely to be attacked. 
To answer (a), we attack three models under additive perturbation and prepending perturbation in \S~\ref{sec:exp_add} and \S~\ref{sec:exp_pp} respectively. Contrasting results in \S~\ref{sec:exp_add} and \S~\ref{sec:exp_pp} answers question (b). We vary different transcripts in \S~\ref{sec:exp_tgt} for question (c). Lastly we experiment with different max-nrom constraints in \S~\ref{sec:exp_norm} and discuss baselines in \S~\ref{sec:exp_baseline}.

\subsection{Attack with additive perturbation} \label{sec:exp_add}
\begin {figure}[tbp]
% \captionsetup{farskip=0pt}% <--- no gap at the top
\centering
\begin{tabular}{cc}
    % \hspace{-0.2 in}
    \includegraphics[width=0.17\textwidth]{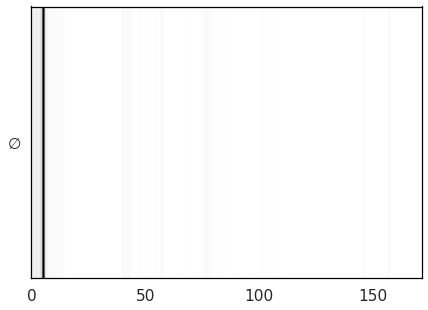}
    & % \hspace{-0.28 in}
    \includegraphics[width=0.205\textwidth]{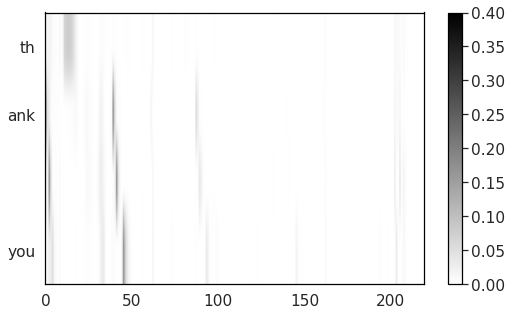} \\
    (a)  $y'=$ $\varnothing$ & 
    (b)  $y'=$ ``thank you''
\end{tabular}
\caption{Heatmap of the attention probability of a perturbed input example on LAS. The x-axis is the encoder frame, and the y-axis is the output token. The attention only focuses on the beginning of the utterance. It quickly emits the target $y'$ and EOS, and skips the rest of the frames. The attention is the culprit for LAS vulnerability under adversarial perturbations.}
\label{fig:attn}
\vspace{-0.1 in}
\end{figure}

Table~\ref{tab:additive} shows the attack results of 4 second additive perturbation across three models. For both targets empty string ($\varnothing$) and ``thank you'', LAS model achieves almost 100\% success with relatively small perturbation. The failure of LAS comes from that the attention is fooled to only focus on the beginning of the speech where the perturbation is applied and ignores the rest of the speech, thus generates the wrong target transcript. See Fig.~\ref{fig:attn} for an illustration.
\begin{table}[t]
    \centering
    \caption{Attack results of 4 seconds additive perturbation on different models. LAS model can be attacked with almost 100\% accuracy. RNN-T and CTC models are more robust.}    \label{tab:additive}
    \begin{tabular}{|c|c|c|c|}
    \hline
    target $y'$ & model  & dB $\downarrow$ & success (\%) $\uparrow$ \\ \hline
    \multirow{3}{*}{$\varnothing$} 
    & LAS &  -7.69 / -6.82 & 99.35 / 99.66 \\ 
    & RNN-T & 6.00 / 6.87 & 14.66 / 19.02 \\ 
    & CTC & 6.00 / 6.87 & 3.28 / 4.80 \\ \hline
     \multirow{3}{*}{``thank you''}
    & LAS &  -0.55 / 0.32 & 98.74 / 99.52 \\ 
    & RNN-T &  6.00 / 6.87 & 1.85 / 2.75  \\ 
    & CTC & 6.00 / 6.87 & 0.00 / 0.00\\ \hline 
    \end{tabular}
    \vspace{-0.1 in}
\end{table}

\begin {figure}[tbp]
% \captionsetup{farskip=0pt}% <--- no gap at the top
\centering
\begin{tabular}{cc}
    \vspace{-0.06 in}
    \hspace{-0.2 in}
    \includegraphics[width=0.25\textwidth]{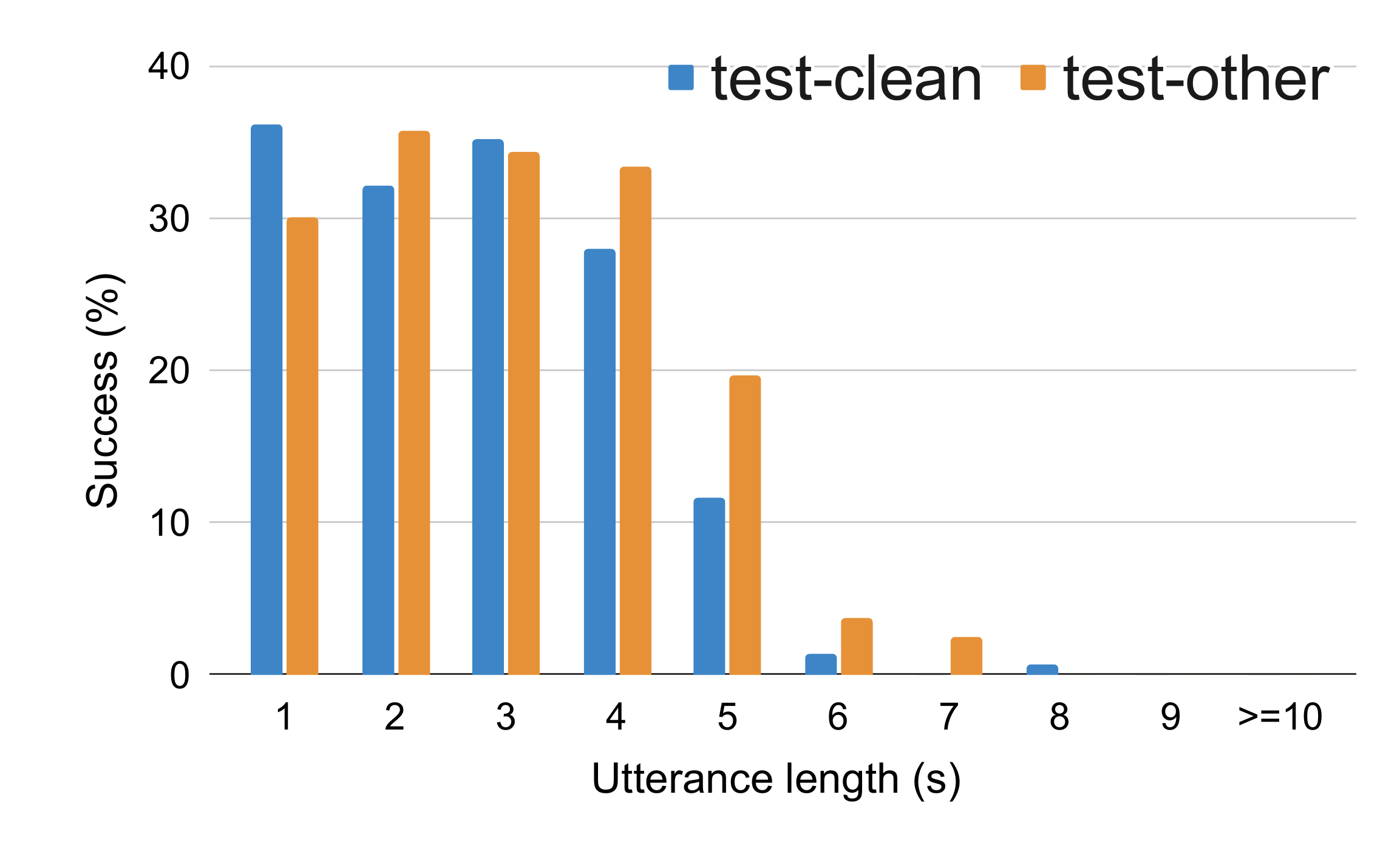}
    &\hspace{-0.28 in}
    \includegraphics[width=0.25\textwidth]{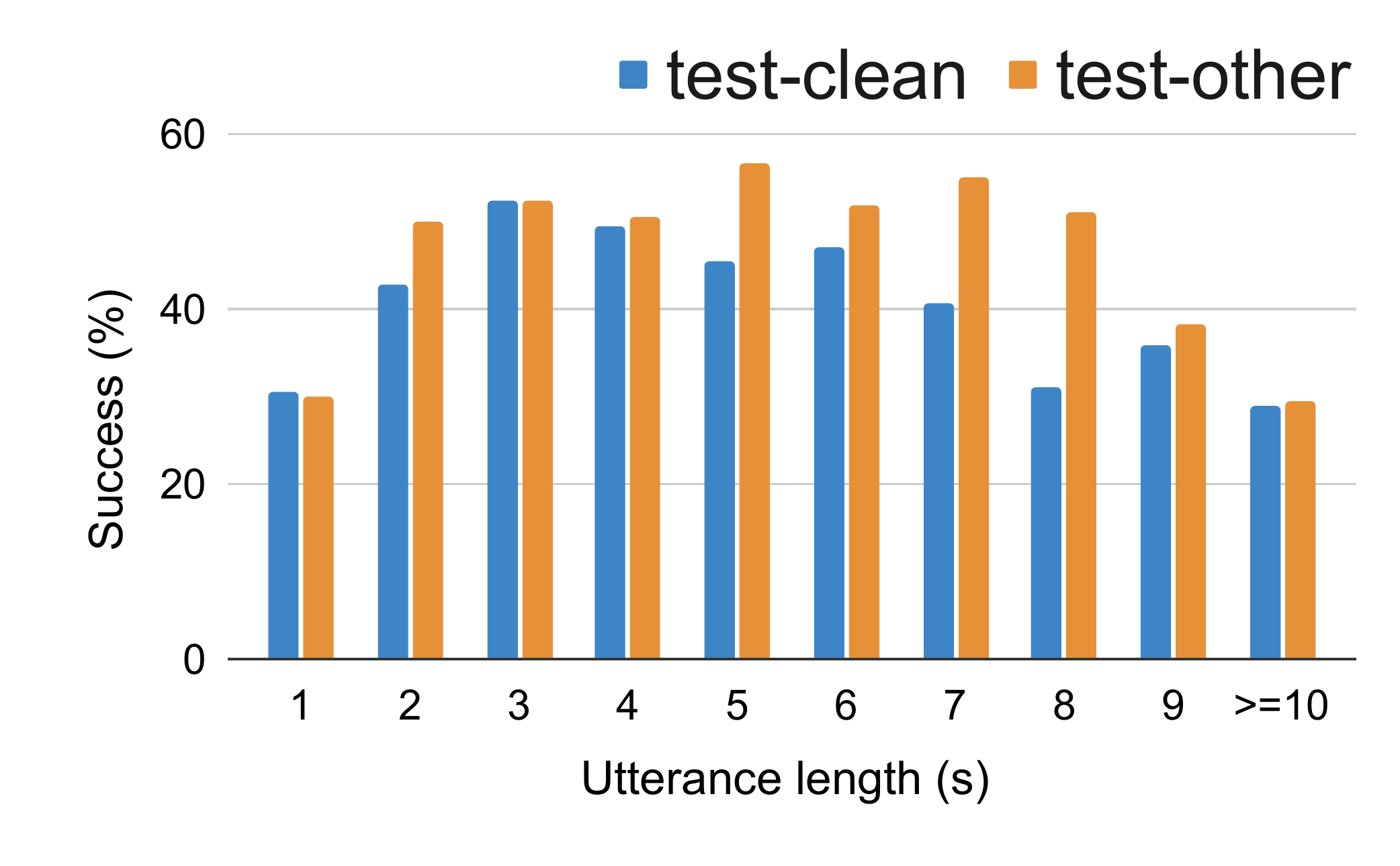} \\
    \hspace{-0.1 in} (a)  additive perturbation & 
    \hspace{-0.15 in} (b)  prepending perturbation
\end{tabular}
\vspace{-0.1 in}
\caption{Attack success rate across different length of utterances. The model is RNN-T and the target is $y'=\varnothing$. The perturbations are 4 seconds long. For additive perturbation, the success rate drops significantly when the length of the utterance is longer than the perturbation. For prepending perturbation, the success rate is relatively uniform across different lengths.} \label{fig:succ_len}
\vspace{-0.1 in}
\end{figure}

However, when the decoder is time-synchronous, \ie RNN-T and CTC, the attack is less successful. Despite a very large perturbation magnitude, the success rate of RNN-T is less than 20\% for empty target $\varnothing$, and less than 3\% for ``thank you'' target. The success rate of CTC is less than 6\% for $\varnothing$, and 0\% for ``thank you''. The reason is that it is challenging to attack arbitrary length utterance by adding a fixed length perturbation. When the utterance is longer than the added perturbation, it is hard to alter the output of those clean speech frames for RNN-T and CTC decoders. To verify this, we plot the attack success rate across different utterance lengths for RNN-T model in Fig.~\ref{fig:succ_len} (a). The target is $\varnothing$. We can see that the success rate drops significantly when $\vx$ is longer than 4 seconds, which is the length of the perturbation.

\begin{table*}[t]
    \centering
    \caption{Example ASR output of untargeted attack of 4 seconds additive perturbation on LAS. The bold part shows that untargeted attack can keep the same text from ground truth in the prediction, despite high WER. }
    \begin{tabular}{rrccccc}
   \hline
    \bf{Hyp:} & 
    \makecell[r]{and but martha and his \\
    wife's rather planted with} & \bf{himself}
    & for the first of the & \bf{better} & deserts of & \bf{anything else he said}\\         \hline
     \bf{Ref:} & {i really liked that account of} & \bf{himself}& & \bf{better} & than & \bf{anything else he said} \\         \hline
      \bf{Err:} & \multicolumn{1}{c}{\emph{ins/del}}  & \emph{correct} & \emph{ins} & \emph{correct} & \multicolumn{1}{c}{\emph{sub}} & \emph{correct} \\ \hline
    \end{tabular}
    \label{tab:untgt}
    \vspace{-0.1 in}
\end{table*}

\subsection{Attack with prepending perturbation} \label{sec:exp_pp}
Before we present the attack results, please see the last column of Table~\ref{tab:wer} for the WER of a simple baseline, which pads 4 seconds of silence (numerical 0) before the utterance. We can see that CTC model is stable with respect to the silence padding. There are some WER degration on RNN-T and LAS, but silence padding does not fail the ASR for most of the time.

In Table~\ref{tab:prepend}, we show the attack results of prepending perturbation. Again for both targets $\varnothing$ and ``thank you'', LAS model achieves almost 100\% success. The perturbation magnitude of prepending perturbation is smaller than that of additive perturbation, when we compare the first row of Table~\ref{tab:prepend} and Table~\ref{tab:additive}.

On RNN-T model, prepending perturbation achieves higher success rate with smaller perturbation magnitude, compared to additive perturbation in Table~\ref{tab:additive}. And if we plot out the attack success rate across different utterance lengths, it is almost uniform across different lengths as seen in Fig.~\ref{fig:succ_len} (b). We hypothesize that prepending perturbation leverages a different failure mode than the additive instability in RNN-T model. By prepending the perturbation, we let the decoder runs into bad RNN states and fail the alignment in RNN-T. It makes the attack agnostic to the length of the utterance. 

\begin{table}[t]
    \centering
    \caption{Attack results of 4 seconds prepending perturbation on different models. Prepending perturbation is more effective than additive perturbation in terms of success rate and distortion on both LAS and RNN-T.}
    \begin{tabular}{|c|c|c|c|}
    \hline
    target $y'$ & model & dB $\downarrow$ & success (\%) $\uparrow$ \\ \hline
    \multirow{3}{*}{$\varnothing$} 
    & LAS &  -30.86 / -29.99 & 99.54 / 99.39 \\
    & RNN-T & -3.51 / -2.64 & 41.18 / 47.43 \\ 
    & CTC &  -4.72 / -3.85 & 1.26 / 0.99 \\ \hline
     \multirow{3}{*}{``thank you''}
    & LAS &  -23.52 / -22.65 & 99.92 / 99.93\\
    & RNN-T &  -0.24 / 0.63 & 35.65 / 37.80 \\ 
    &  CTC &  -0.27 / 0.60  & 0.00 / 0.07 \\ \hline
    \end{tabular}
    \label{tab:prepend}
    \vspace{-0.1 in}
\end{table}
Lastly, CTC model is robust to the prepending attack due to the conditional independence modeling assumption.

\subsection{Attack results on more mis-transcript targets} \label{sec:exp_tgt}
In Table~\ref{tab:target}, we discuss the results of different targets $y'$ with prepending perturbation on LAS. We choose LAS as the model and prepending perturbation because they are relatively ``easier'' to learn, according to results from Table~\ref{tab:additive} and~\ref{tab:prepend}. As can be seen from Table~\ref{tab:target}, the general trend is that the success rate drops when $y'$ gets longer, like `'to be or no to be'', and when $y'$ consists of less frequent words, like ``carpe diem'' in Latin. It is worth mentioning that the learnt perturbation contains phonetic similar sound as the ones of the target $y'$.
\begin{table}[t]
    \centering
    \caption{Comparing different target with prepending perturbation on LAS. The shorter the transcript, and the more frequent the words, the easier the attack can succeed.} \label{tab:target}
    \begin{tabular}{|c|c|c|}
    \hline
    target $y'$ & dB $\downarrow$ & success (\%) $\uparrow$ \\ \hline
    $\varnothing$ &    -30.86 / -29.99 & 99.54 / 99.39  \\
    ``hello'' &    -23.58 / -22.71 & 99.77 / 100.00 \\ 
    ``thank you'' &     -23.52 / -22.65 & 99.92 / 99.93 \\
    ``carepe diem'' &   -11.25 / -10.38 & 95.69 / 92.04 \\ 
    ``to be or not to be'' &  -18.11 / -17.24 & 66.87 / 71.76 \\ \hline
    \end{tabular}
    \vspace{-0.2 in}
\end{table}

When the target $y'$ is long, it is hard to be attacked. For example, ``to be or no to be that is the question'' achieves 1.57\% and 2.01\% attack accuracy on test-clean and test-other sets, and ``the gods had condemned sisyphus to ceaselessly rolling a rock to the top of a mountain whence the stone would fall back of its own weight'' achieves 0\% accuracy on both test sets. Similar observation is discussed for per-utterance targeted attack~\cite{carlini2018audio}.

\subsection{Attack results under different max-norm constraints} \label{sec:exp_norm}
Lastly, we vary the max norm constraint $\epsilon$ and report the attack results. In Table~\ref{tab:norm}, we can see that the success rate drops when the $\epsilon$ gets smaller.
\begin{table}[t]
    \centering
    \caption{Attack results under different max-norm constraints. The perturbation is 4 second prepending perturbation, and the model is RNN-T. The success rate drops when  $\epsilon$ become smaller.}\label{tab:norm}
    \begin{tabular}{|c|c|c|}
    \hline
    $\|\vdelta\|_\infty$ & dB $\downarrow$ & success (\%) $\uparrow$ \\ \hline
    $2^{15}$ &   -3.51 / -2.64 & 41.18 / 47.43 \\
    4000 & -12.25 / -11.38   & 28.63 / 31.30 \\ 
    2000 &  -18.25 / -17.38  & 3.32 / 3.74 \\ \hline
    \end{tabular}
    \vspace{-0.1 in}
\end{table}

\subsection{Comparing with other adversarial attacks} \label{sec:exp_baseline}
In this section, we briefly compare our study with two existing works~\cite{qin2019imperceptible,neekhara2019universal}.
Note that this paper studies a different problem from theirs, since the perturbation is input-agnostic as well as targeted. In addition, we experiment with both additive and prepending perturbations, while~\cite{qin2019imperceptible,neekhara2019universal} are limited to additive perturbations. However, it would be intuitive to compare different attacks and discuss why their method cannot solve the universal targeted perturbation problem. 

\noindent \textbf{Targeted per-utterance perturbation.} ~\cite{qin2019imperceptible} solves
$ \displaystyle
\min_{\vdelta} \ell(\sT(\vdelta, \vx), y'), \text{s.t. } \|\vdelta\|_{\infty} < \epsilon,  \label{eq:per-utt}
$
for a $(\vx, y')$ utterance-text pair\footnote{For simplicity, here we omit the irrelevant optimization over the imperceptibility and robustness of the perturbation.}. We randomly pick 5 utterances from the Librispeech100h clean training set to learn 5 additive perturbations on LAS, which is the same length as its corresponding training utterance. The target is empty string.  However, for the perturbation trained from a single utterance, the success rates on test-clean is $0.7\%\pm0.7\%$ and test-other $2.7\%\pm3.7\%$, averaging over the 5 perturbations. It shows that the perturbation optimized for one utterance can hardly generalize to other unseen utterances.

\noindent \textbf{Untargeted universal perturbation.} ~\cite{neekhara2019universal} studies universal perturbation for \emph{untargeted} attack. 
The loss is
$\displaystyle \max_{\vdelta} \sum_{x\in \mathcal{D}} \ell(\sT(\vdelta, \vx), y), \text{s.t. } \|\vdelta\|_{\infty} < \epsilon,$
where $y$ is the ground-truth transcription. The goal is to find a perturbation that maximally deviates the predictions from the ground-truth, but the error is not specified. We implement their universal attack on Librispeech960h. The WERs in percentage (\%) on test-clean/test-other for 3 models are: LAS 76.0/100.5, RNN-T 87.5/100.6, and  CTC 106.1/121.2. Table.~\ref{tab:untgt} shows an example of such attack. Despite high WER, the predictions on the untargeted perturbation  oftentimes preserves some of the ground truth's texts. In contrast, our targeted attack will keep no words from the original ground-truth, and hence is more surprising from the optimization perspective.   

%!TEX root = main.tex
\section{Conclusion} \label{sec:conclude}
In this work, we study the problem of targeted universal adversarial perturbations. We compare the performance of both additive and prepending perturbations against three state-of-the-art e2e ASR models, and suggest that targeted universal attacks exist for both LAS and RNN-T, but not for CTC models. Our future work is to develop more robust e2e models based on the lessons learnt from this study. 

\section{Acknowledgements}
We are very thankful for our colleagues Hasim Sak, Oren Litvin, Shuai Shao, Chung-Cheng Chiu, Chao Zhang, Arun Narayanan,  Yongqiang Wang, Yash Sheth, Sean Campbell, Rohit Prabhavalkar, Pedro Moreno Mengibar and Trevor Strohman for their help and suggestions.

\bibliographystyle{IEEEtraN}

\bibliography{main}

\end{document}